\documentclass[aps,prb,twocolumn,floatfix]{revtex4}
\usepackage{graphicx}

\begin{document}
\title{
The Friedel oscillations in the presence of transport currents
}
\author{Anna Gorczyca}
\author{Maciej M. Ma{\'s}ka}
\author{Marcin Mierzejewski}
\affiliation{
Department of Theoretical Physics, Institute of Physics,
University of Silesia, 40-007 Katowice,
Poland}

\begin{abstract}
We investigate the Friedel oscillations in a nanowire coupled to two
macroscopic electrodes of different potentials. We show that the wave--length of the density oscillations
monotonically increases with the bias voltage, whereas the amplitude and the spatial decay exponent of the
oscillations remain intact.  Using the nonequilibrium Keldysh Green functions, 
we derive an explicit formula that describes voltage dependence of the wave--length of the Friedel oscillations. 
\end{abstract}
\pacs \ 73.63.Nm, 73.63.-b, 71.45.Lr
\maketitle

\section{Introduction}
Transport properties of nanosystems, e.g., nanowires or single molecules, have 
recently been receiving significant attention 
mainly due to their possible application in future electronic devices. \cite{tr1,tr2,tr3,tr4}
These properties strongly differ from those of macroscopic conductors.
The most important obstacle in theoretical investigations 
of the transport phenomena originates from the coupling 
between nanosystem and macroscopic leads. 
Because of this coupling, analysis of the electron correlations
is more difficult than in the equilibrium case.

In nanosystems the charge carriers are usually distributed inhomogeneously. 
There exist several reasons for such an inhomogeneity:

{\em (i)} First, it may originate from a spatial confinement. \cite{confinement} 
In analogy to the case of a quantum well, one may expect that due to a 
small size of nanosystems, electrons are inhomogeneously distributed.
In particular, recent scanning tunneling spectroscopy
has shown the presence of the electronic standing waves at the end of a 
single--wall carbon nanotube.\cite{nanotube}  

{\em (ii)} Additionally, in the transport phenomena it may originate from 
the applied voltage. \cite{bulka,emberly,tr1} In this case the  system properties are determined 
by the chemical potentials of the left and right electrodes.
Different values of these potentials may lead to an inhomogeneous 
distribution as well. 

{\em (iii)} Similarly to the macroscopic case, inhomogeneous charge distribution in nanosystems
should occur in the presence of impurities.\cite{friedel}   

{\em (iv)} Nanowires or molecular wires represent quasi--one--dimensional 
conductors. Therefore, phenomena typical for low dimensional systems, 
e.g., charge density waves may occur as well. \cite{markovic, mantel, ringland, oxman, krive} 
Recently it has been shown that
the charge density waves are strongly modified by the bias voltage. \cite{my1}
Apart from the low--voltage regime, they are incommensurate and the corresponding 
wave vector decreases discontinuously with the increase of the bias voltage.

In this paper we focus on the impurity--induced inhomogeneities. It is known that
an impurity in the electron gas produces local changes of the carrier concentration, 
known as the Friedel oscillations \cite{friedel} that asymptotically decay with the distance 
from the impurity. 
The most of recent theoretical investigations of the Friedel oscillations concerned the
influence of the electronic correlations, that is of crucial importance 
in one--dimensional systems.\cite{eckern,cohen,eggert,rommer,weiss}
It has been shown that correlations suppress the decay of the density oscillations.\cite{cohen,rommer}
It is interesting that these oscillations give information about the impurities\cite{rommer} as well
as the electron--electron interaction in Luttinger liquid systems.\cite{eggert,weiss}             
In macroscopic systems, the Friedel oscillations are closely related to the singularity in the response function 
for wave--vectors close to 2$k_F$, where $k_F$ is the Fermi wave--vector.
If the nanosystem is isolated (or more generally, is in equilibrium), $k_F$ is a well defined quantity.
However, in the transport experiments the nanosystem is coupled to two macroscopic leads 
with different Fermi levels and the difference between these Fermi energies 
increases with the bias voltage. Therefore, the meaning of $k_F$ is ambiguous.
Since the properties of a nanosystem are determined by the chemical potentials of the
left and right electrodes, one may expect that the Friedel oscillations should depend
on the voltage as well. In this paper we analyze this dependence using the formalism of the
nonequilibrium Keldysh Green functions.
In particular, we derive an explicit formula for the voltage dependence of the wave--length
of the Friedel oscillations.

The paper is organized as follows: In Section II we discuss a microscopic model and details
of calculations. Numerical results are presented in Section III. 
Approximate analytical formulas are derived in Section IV. The last
section contains a discussion and concluding remarks.

\section{Model and the calculations scheme}
We investigate a one--dimensional nanowire with its ends coupled to macroscopic leads. The system
under consideration is described by the Hamiltonian 
\begin{equation}
H=H_{\rm el} + H_{\rm nano} + H_{\rm nano-el},\label{ham}
\end{equation}
where $H_{\rm el}$, $H_{\rm nano}$ and $H_{\rm nano-el}$ describe leads,
nanowire, and the coupling between the wire and leads, respectively.                          
We assume that electrodes are described by the free electron gas, with a wide energy band: 
\begin{equation}
H_{\rm el}=\sum_{{\bf k},\sigma,\alpha} \left(\varepsilon_{{\bf k},\alpha} 
-\mu_{\alpha}\right)
c^\dagger_{{\bf k}\sigma\alpha} c_{{\bf k}\sigma\alpha},
\end{equation}
where $\mu_{\alpha}$ is the chemical potential and $\alpha \in$ \{L,R\} indicates the left or
right electrode. $\mu_{L}-\mu_{R}=eV$, with $V$ being the bias voltage. 
 $c^\dagger_{{\bf k}\sigma\alpha} $ creates an electron with
momentum
${\bf k}$ and spin $\sigma$ in the electrode $\alpha$. 
The  Hamiltonian of the  nanosystem is given by
\begin{equation}
H_{\rm nano}= -t \sum_{\langle ij\rangle\sigma} d^\dagger_{i\sigma}d_{j\sigma}
+U \sum_{\sigma} n_{l\sigma}. \label{H_nano}
\end{equation}
Here, $d^\dagger_{i\sigma}$ creates an electron with spin $\sigma$
at site $i$ of the  nanosystem, $n_{i\sigma}=d^\dagger_{i\sigma}d_{i\sigma}$ and
$U$ is the impurity potential. We have assumed a single impurity localized at site $l$.
The coupling between the nanowire and the leads is given by:                                  \begin{equation}
H_{\rm nano-el}=\sum_{{\bf k},i,\alpha,\sigma} \left(
g_{{\bf k},i,\alpha} c^\dagger_{{\bf k}\sigma\alpha }d_{i\sigma}+{\rm H.c.}\right).
\end{equation}
In the following, we assume that the matrix elements $g_{{\bf k},i,\alpha} $ are 
nonzero only for the edge atoms of the nanowire.

The electron distribution has been determined with the help of the
nonequilibrium Keldysh Green functions. Here, we follow
the procedure used by Kostyrko and  Bu{\l}ka in Ref. \onlinecite{bulka}.
In particular, the local carrier density is expressed by the lesser
Green function,
\begin{equation}
\langle d^\dagger_{i\sigma} d_{i\sigma} \rangle
=\frac{1}{2 \pi i} \int {\rm d} \omega \:  G^{<}_{i\sigma,i\sigma} (\omega).
\label{gless1}
\end{equation}
This quantity, in turn, is determined by the retarded and advanced Green functions
in the  following way:    
\begin{equation}
\hat{G}^{<}(\omega)=i \sum_{\alpha \in \{L,R\}}
\hat{G}^r(\omega)
\hat{\Gamma}_{\alpha}(\omega)\hat{G}^a(\omega)f_{\alpha}(\omega),
\label{gless2}
\end{equation}
where
\begin{equation}
\left[\hat{\Gamma}_{\alpha}(\omega)\right]_{ij}= 2 \pi \sum_{{\bf k}}
g^{*}_{{\bf k},i,\alpha} g_{{\bf k},j,\alpha} \delta(\omega-\varepsilon_{{\bf k},\alpha}),
\end{equation}
and $f_{\alpha}(\omega)$ stands for the Fermi distribution function of the electrode
$\alpha$.
The retarded Green function can be calculated from the following formula:
\begin{equation}
\hat{G}^{r}(\omega)=\left[\omega \hat{I}-\hat{H}-\hat{\Sigma}^{r}(\omega)\right]^{-1}, 
\label{gret}
\end{equation}
where $\hat{H}$ consists of the matrix elements of $H_{\rm nano}$, i.e.,
$
H_{mn}=-t \delta_{m,n\pm 1}+U \delta_{m,l} \delta_{n,l}
$
and the 
 retarded self--energy is determined by the coupling between the nanowire and the leads
\begin{equation}
\hat{\Sigma}^{r}(\omega)= \frac{1}{2}\sum_{\alpha \in \{L,R\}} \left[
\frac{1}{\pi}P \int d \Omega \frac{ \hat{\Gamma}_{\alpha}(\Omega)}{\omega-\Omega}
-i \hat{\Gamma}_{\alpha}(\omega) \right]. \label{sigmar}
\end{equation}

\section{Numerical Results}

We have solved numerically the system of Eqs. (5-9) for nanowires consisting
of up to  $N=129$ 
lattice sites, with a single impurity in the middle of the wire.
The only non--vanishing elements of $\hat{\Gamma}$'s have been assumed 
to be frequency independent
$\left[\hat{\Gamma}_{\rm L}(\omega)\right]_{11}=\left[\hat{\Gamma}_{\rm R}(\omega)\right]_{NN}
=\Gamma_0$, where the sites in the chain are enumerated from 1 to $N$. We have taken the 
nearest neighbor hopping integral $t$ as an energy unit and assumed the coupling between the 
nanosystem and the leads as $\Gamma_0=0.1$. The temperature of both the leads is $k_BT=0.01$.
Figure 1 shows the spatial distribution of electrons in the nanowire for $U=2$ and various values of the
bias voltage. One can see strong oscillations in the vicinity of the impurity.
However, due to the coupling to the leads the electron distribution visibly differs from
the standard Friedel oscillations:
\begin{equation}
n(x)=\bar{n}+A \cos \left(Q x+\eta\right)/x^\delta,
\label{fridel}
\end{equation}
where the wave--vector $Q=2k_F$ and the parameters $\eta$ and $\delta$ 
depend on the interaction.
In our case, the actual value of $Q$ has been obtained 
from the fast Fourier transform  of the electron distribution $n(x)$.

\begin{figure}
\includegraphics[width=8cm]{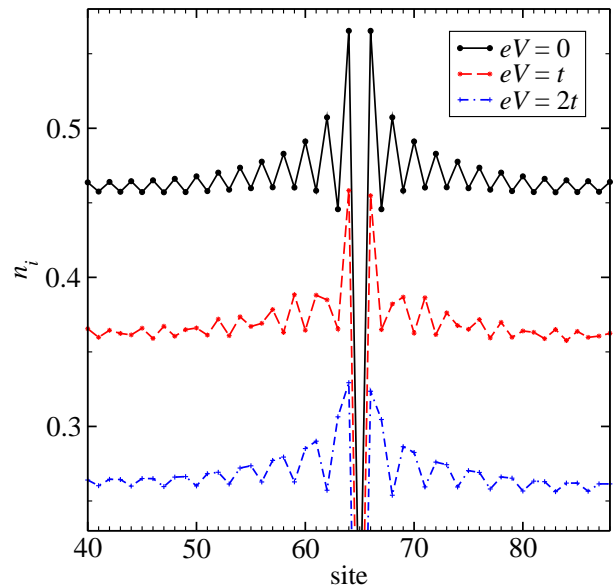}
\caption{(Color online) 
Occupation of sites in the vicinity of the impurity for a 129--site nanowire.
The bias voltage is explicitly indicated in the legend. 
For the sake of clarity,  the curves for $eV=t$ and $eV=2t$ have been shifted downward
by $0.1$ and $0.2$, respectively.
}
\label{fig1}
\end{figure}

Figure 2 shows the voltage dependence of the vector $Q$.
In the equilibrium case ($V=0$)  $Q=\pi$  (with the lattice constant $a=1$), so the charge
oscillations are commensurate with the lattice.  
Since in the half--filled case $k_F=\pi/2$, this wave--vector remains in agreement
with the standard relation $Q=2k_F$.
However, when the bias voltage is switched on, 
the situation changes and the oscillations are in  general no longer commensurate. 
Moreover, one can see a  strong dependence of $Q$ on the applied voltage.
$Q$ is a monotonically decreasing function of voltage 
and vanishes for a sufficiently large $V$.   
Similar situation occurs in the transport
phenomena through one--dimensional charge density wave systems. \cite{my1}
Our numerical results indicate that the obtained $Q(V)$ dependence can be very accurately described by 
a  formula
\begin{equation}
eV= 4t\cos(Q/2). \label{fit}
\end{equation}
The surprising simplicity of Eq. (\ref{fit}) is very suggestive.
In the following Section we present an approximate analytical 
approach that explains such a form of $Q(V)$. It is applicable for arbitrary 
tight--binding Hamiltonian of noninteracting electrons and holds true in
 a wide range 
of the coupling strength $\Gamma_0$. The numerical results presented in Figure 2
allow us to test the applied approximations.

\begin{figure}
\includegraphics[width=8cm]{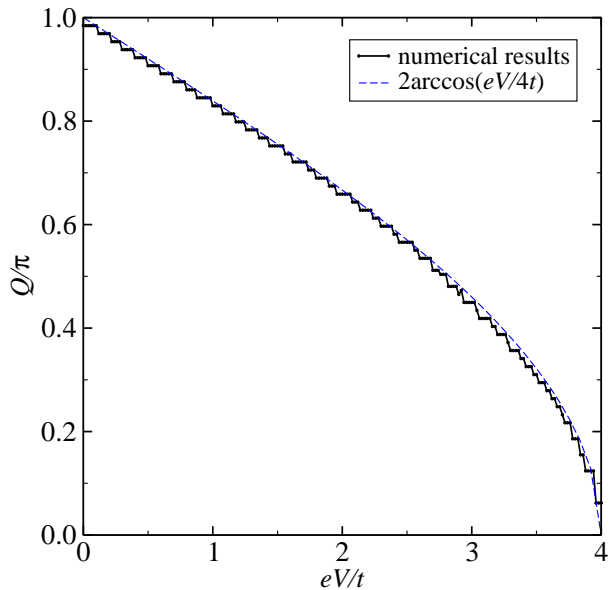}
\caption{(Color online) The wave--vector $Q$ of the charge oscillations 
calculated by means of the fast Fourier transform of the numerical solution
of Eqs. (5-9) (solid line). The dashed line shows the fit given by Eq. \ref{fit}. The numerical 
results have been obtained for the same parameters as in Fig. 1 The discreteness of $Q(V)$ comes from
the finite number of the lattice sites.}
\label{fig2}
\end{figure}

\section{Analytical discussion}

In the equilibrium case, the eigenstates of an isolated systems with
{\em periodic boundary conditions} (pbc) are built out of plane waves. 
The Friedel oscillations are related to the 
maximum in the response function defined as a retarded Green function:
\begin{equation}
\chi(Q,\omega)=-\langle\langle \hat{\rho}(Q) \mid  \hat{\rho}^\dagger(Q)
\rangle\rangle, \label{sus}
\end{equation}
calculated for $U=0$. Here,
\begin{eqnarray}
\hat{\rho}(Q) &=& \sum_{i,\sigma} \exp (i Q R_i) d^{\dagger}_{i\sigma} d_{i \sigma} \nonumber \\
&=&  \sum_{{k},\sigma} d^{\dagger}_{{k}+Q\sigma} d_{{k}\sigma}, 
\end{eqnarray}
where the summation is carried out over all momenta $k$.  
In the following we demonstrate that this quantity helps one to explain the dependence
$Q=Q(V)$ also in the nonequilibrium case.

When the nanosystem is connected to macroscopic leads, the pbc 
become inappropriate since they do not reflect the geometry 
of the  experimental setup. Then, the choice of
{\em open boundary conditions} (obc) seems to be more appropriate. 
For $U=0$, the Hamiltonian (\ref{H_nano}) with obc can be diagonalized with
the help of the unitary transformation,\cite{mylast}     
\begin{equation}
d_{i\sigma}^{\dagger}=\sqrt{\frac{2}{N+1}}\sum_{k}\sin(k R_{i})d_{k\sigma}^{\dagger},
\end{equation}
where the wave--vectors $k$ take on the following values
\begin{equation}
k=\frac{\pi}{N+1}, \frac{2\pi}{N+1}, \dots, \frac{N\pi}{N+1}.
\end{equation} 
The specific form of this transformation accounts
for vanishing of the one--electron wave functions at the edges of the nanosystem.
In this representation one gets
\begin{equation}
H_{\mathrm{nano}}=\sum_{k \sigma}\epsilon_{k} d_{k\sigma}^\dagger d_{k\sigma},
\end{equation}
where
\begin{equation}
\epsilon_{k}=-2t\cos(k).
\end{equation}
Although, the dispersion relation is exactly the same as for pbc, 
the values of $k$ belong to $(0,\pi)$ instead of the 1st Brillouin zone $(-\pi,\pi)$. 
One can apply the above transformation also to the remaining terms in the Hamiltonian
(\ref{ham}) and repeat calculations presented in Sec. II.
The resulting equations have the same structure as Eqs. (5-9) with the real space
variables $i$ replaced by the wave--vectors $k$. In the new representation, 
the Hamiltonian matrix $\hat{H}$ is diagonal, however, the matrices $\hat{\Gamma}_{\alpha}$ take on
much more complicated form:     
\begin{equation}
\left[\hat{\Gamma}_{\alpha}(\omega)\right]_{kp}= \frac{2}{N+1} \sum_{ij}
\left[\hat{\Gamma}_{\alpha}(\omega)\right]_{ij}\sin (k R_{i})\sin (p R_{j}).
\label{gamak}
\end{equation}

In order to analyze the Friedel oscillations we investigate the correlation
function given by Eq. (\ref{sus}) with
\begin{eqnarray}
\hat{\rho}(Q)&=& \sum_{i \sigma} \cos(Q R_i) d^{\dagger}_{i \sigma} d_{i \sigma} \nonumber \\
&=&\sum_{kp\sigma} {\mathcal{B}}_{Q}(k,p) d_{k\sigma}^{\dagger}d_{p\sigma},
\end{eqnarray}
where
\begin{equation}
\mathcal{B}_{Q}(k,p)=\frac{1}{2}\left(\delta_{p,k-Q}-\delta_{p,Q-k}+\delta_{p,k+Q} 
-\delta_{p,2\pi -(k+Q)} \right) \label{B_matrix}.
\end{equation}
Equations of motion allow one to calculate the correlation function that  
in the static limit takes on the form
\begin{eqnarray}
\chi(Q,\omega \rightarrow 0)  &=& \sum_{k, p, q, \sigma}\frac{\mathcal{B}_{Q}(k,p)}
{\epsilon_{p}-\epsilon_{k}} 
\left(\mathcal{B}_{Q}^{*}(p,q)\langle d_{k\sigma}^{\dagger}d_{q\sigma}\rangle  \right. 
\nonumber \\
&& \left. -\mathcal{B}_{Q}^{*}(q,k)\langle d_{q\sigma}^{\dagger}d_{p\sigma}\rangle   \right)
+\chi'.
\label{chidok}
\end{eqnarray}
The second term in the above equation, $\chi'$, is proportional to 
$g \langle\langle d^{\dagger} c|d^{\dagger}d \rangle\rangle $ and will be
neglected in the following analysis. Such simplification is justified only for a weak coupling
between the nanosystem and the leads.
In order to demonstrate the validity of this approximation we have calculated numerically
the resulting correlation function for nanowires consisting of 
20 and 40 sites (see Fig. 3). 
The discreteness of the system is clearly visible for short nanowires, whereas for 
large systems the correlation function becomes smoother.  In the latter case 
one can see that $\chi(Q,\omega \rightarrow 0)$ reaches its maximum value for $Q$
given by Eq. (\ref{fit}), what justifies the applied approximation  $\chi' \simeq 0$.  
\begin{figure}
\includegraphics[width=7cm]{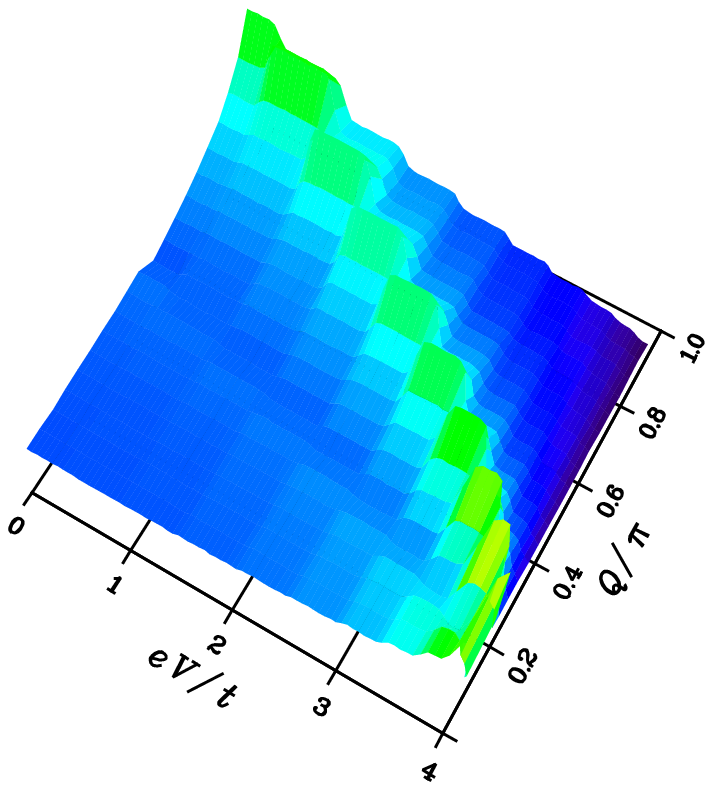}
\includegraphics[width=7cm]{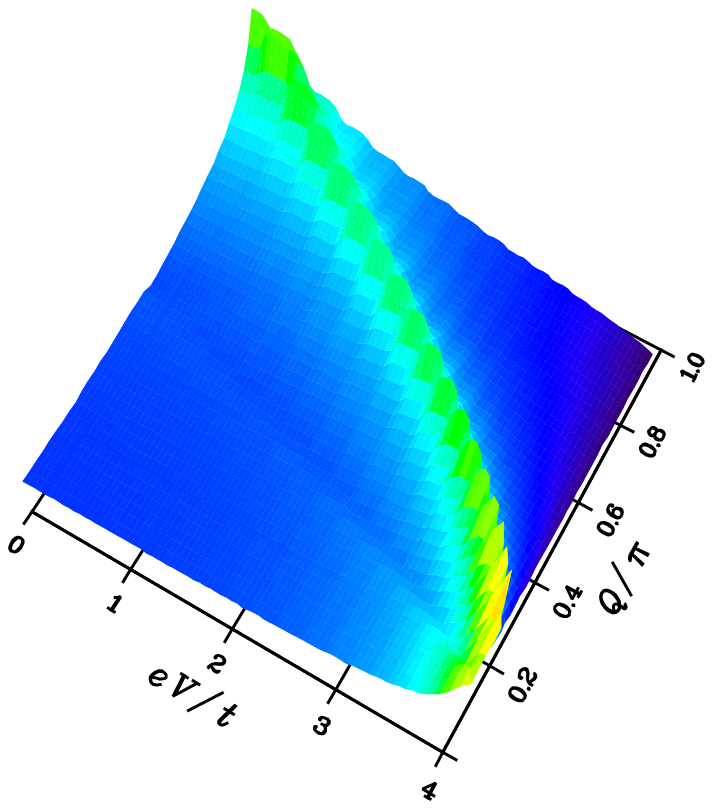}
\caption{(Color online) Correlation function $\chi=\chi(Q,V)$ (Eq. \ref{chidok}) determined numerically for 
20--site (upper panel) and 40--site (lower panel) chain.}
\label{fig3}
\end{figure}

The correlation function given by Eq. (\ref{chidok})
is still too complicated for analytical discussion. The averages in
Eq. (\ref{chidok}) are determined by the lesser Keldysh Green functions that, in turn,
depend on the retarded ones. Because of the coupling between the nanosystem and
the leads, off-diagonal elements of the retarded Green function are nonzero 
[see Eqs. (\ref{sigmar}) and (\ref{gamak})]. In order to proceed with the discussion of
the Friedel oscillations we apply an additional approximation:   
\begin{equation}
\left(\hat{\Gamma}_{\alpha}\right)_{kp} \simeq \delta_{kp} \Gamma'  
\label{przyb}
\end{equation}
Then, all matrices in Eq. (\ref{gless2}) become diagonal. In the next step
we assume $\Gamma' \rightarrow 0$, what allows one to calculate the integral over frequencies in Eq. 
(\ref{gless1}). The resulting correlation function can be expressed as a sum of two Lindhard functions:


\begin{equation}
\chi(Q,\omega\rightarrow 0)=\chi^{L}(Q)+\chi^{R}(Q),
\label{chiprzy1}
\end{equation}
where
\begin{equation}
\chi^{L(R)}(Q)=  \sum_{k, p, \sigma}|\mathcal{B}_{Q}(k,p)|^{2}\frac{f_{L(R)}\left(\epsilon_{k} \right) - f_{L(R)}\left(\epsilon_{p} \right)}{\epsilon_{p} - \epsilon_{k}}.
\label{chiprzy2}
\end{equation}
The Fermi distribution functions of the left and right electrodes read
\begin{equation}
f_{L}\left(\epsilon_{k} \right)=f\left(\epsilon_{k}-\frac{eV}{2}\right),
\;\;
f_{R}\left(\epsilon_{k} \right)=f\left(\epsilon_{k}+\frac{eV}{2}\right),
\end{equation}
with $f(x)=[\exp(x/k_BT)+1]^{-1}$.
The maximum of the response function occurs for such $Q$, that both the Fermi functions in the
numerator in Eq. (\ref{chiprzy2}) vanish simultaneously. It is easy to check that for both 
$\chi^{L}(Q) $ and $\chi^{R}(Q)$ this requirement is equivalent to Eq. (\ref{fit}).
 
At this stage a comment on the approximation given by Eq. (\ref{przyb}) is necessary.
It is a crude and generally inappropriate approximation that strongly affects  
most of the system's properties. In particular, it would strongly modify the 
current--voltage characteristics. However, the 
correlation functions calculated from Eqs. (\ref{chidok}) and (\ref{chiprzy1})
are almost indistinguishable, what {\em a posteriori} 
justifies the use of this approximation for the discussion of charge inhomogeneities.
This surprising result gives some insight into the physical mechanism of the
charge distribution in nanosystems in the presence of transport currents.
This distribution seems to be independent of the details of the coupling between
the nanosystem and the leads however, it is determined by the fact that nanosystem
is connected to two macroscopic particle reservoirs with different chemical potentials.

\section{Concluding remarks}

Using the nonequilibrium Keldysh Green functions 
we have investigated the Friedel oscillations in a nanowire coupled 
to two macroscopic electrodes.  
We have derived a simple formula for the correlation function that determines
the wave vector $Q$ of the oscillations. The approximate analytical expression 
 fits the numerical results obtained from the Fourier transform       
of the electron distribution very accurately.  Our analysis concerns nanosystems described 
by the tight-binding Hamiltonian with the nearest neighbor hopping. However, it
can be straightforwardly extended to account for arbitrary hopping matrix elements.

The above discussion of the Friedel oscillations focuses on the voltage
dependence of the wave--vector $Q$. We have found that the envelope of the 
charge density oscillations is almost bias--voltage independent. It means
that the remaining parameters characterizing the Friedel oscillations, i.e.,
the amplitude $A$ and the spatial decay exponent $\delta$, are determined 
predominantly by the internal properties of the nanowire, whereas the
wave--length of the oscillations depends on the bias--voltage. 
We believe that investigations of the Friedel oscillations in the
transport phenomena should allow one to get insight into many
important parameters of the experimental setup.

\acknowledgements
This work has been supported by
the Polish Ministry of Education and Science under Grant No. 1~P03B~071~30.

\end{document}